\documentclass[superscriptaddress,amssymb,prx,10pt,twocolumn]{revtex4-1}
\usepackage{amsmath}    % need for subequations
\usepackage{graphicx}   % need for figures
\usepackage{verbatim}   % useful for program listings
\usepackage{color}      % use if color is used in text
\usepackage{subfigure}% use for side-by-side figures
\usepackage{gensymb}
\usepackage{textcomp}
\usepackage[bookmarks=false]{hyperref}   % use for hypertext links, including those to external documents and URLs
\usepackage{longtable}
\usepackage{bm}
\usepackage{color}
\usepackage{epstopdf}
\usepackage{physics} % Bra-Ket notation
\usepackage{natbib} % For citations

\hypersetup{
  colorlinks   = true, %Colours links instead of ugly boxes
  urlcolor     = blue, %Colour for external hyperlinks
  linkcolor    = blue, %Colour of internal links
  citecolor   = blue %Colour of citations
}

\graphicspath{{Figures//}}

\newcommand{\oplsco}{\ensuremath{\mathrm{La_{1.85}Sr_{0.15}CuO_4}}}
\newcommand{\lco}{\ensuremath{\mathrm{La_2CuO_4}}}
\newcommand{\slao}{\ensuremath{\mathrm{SrLaAlO_4}}}
\newcommand{\lcmo}{\ensuremath{\mathrm{La_{2/3}Ca_{1/3}MnO_3}}}
\newcommand{\rs}{\ensuremath{\mathrm{R_{\square}}}}

\begin{document}
\title{\textbf{Superconductivity and charge carrier localization in ultrathin  $\mathbf{{La_{1.85}Sr_{0.15}CuO_4}/{La_2CuO_4}}$ bilayers}} % Force line breaks with \\

\author{K. Sen}
\email{kaushik.sen@kit.edu}
\altaffiliation[Present address: ]{Institut f\"ur Festk\"orperphysik, Karlsruher Institut f\"ur Technologie, D-76344,
Eggenstein-Leopoldshafen, Germany
}
\author{P. Marsik}
\author{S. Das}
\altaffiliation[Present address: ]{Department of Physics and Astronomy, Seoul National University (SNU), Seoul 08826, Republic of Korea}
\author{E. Perret}
\author{R. de Andres Prada}
\affiliation{University of Fribourg, Department of Physics and Fribourg Center for Nanomaterials, Chemin du Mus\'ee 3, CH-1700, Fribourg, Switzerland}
\author{A. Alberca}
\affiliation{University of Fribourg, Department of Physics and Fribourg Center for Nanomaterials, Chemin du Mus\'ee 3, CH-1700, Fribourg, Switzerland}
\affiliation{Swiss Light Source, Paul Scherrer Institut, CH-5232, Villigen PSI, Switzerland}
\author{N. Bi\v{s}kup}
\affiliation{Departamento de F\'{i}sica Aplicada III, Instituto Pluridisciplinar, Universidad Complutense de Madrid, Madrid 28010, Spain}
\affiliation{Centro Nacional de Microscop\'{i}a Electr\'{o}nica, Universidad Complutense de Madrid, Madrid 28040, Spain}
\author{M. Varela}
\affiliation{Departamento de F\'{i}sica Aplicada III, Instituto Pluridisciplinar, Universidad Complutense de Madrid, Madrid 28010, Spain}
\author{C. Bernhard}
\email{christian.bernhard@unifr.ch}
\affiliation{University of Fribourg, Department of Physics and Fribourg Center for Nanomaterials, Chemin du Mus\'ee 3, CH-1700, Fribourg, Switzerland}

%% The bibliography style
%\bibliographystyle{naturemag}

\begin{abstract}
{\oplsco}/{\lco} (LSCO15/LCO) bilayers with a precisely controlled thickness of N unit cells (UCs) of the former and M UCs of the latter ([LSCO15\_N/LCO\_M]) were grown on (001)-oriented {\slao} (SLAO) substrates with pulsed laser deposition (PLD). X-ray diffraction and reciprocal space map (RSM) studies confirmed the epitaxial growth of the bilayers and showed that a [LSCO15\_2/LCO\_2] bilayer is fully strained, whereas a [LSCO15\_2/LCO\_7] bilayer is already partially relaxed. The \textit{in situ} monitoring of the growth with reflection high energy electron diffraction (RHEED) revealed that the gas environment during deposition has a surprisingly strong effect on the growth mode and thus on the amount of disorder in the first UC of LSCO15 (or the first two monolayers of LSCO15 containing one $\mathrm{CuO_2}$ plane each). For samples grown in pure $\mathrm{N_2O}$ gas (growth type-B), the first LSCO15 UC next to the SLAO substrate is strongly disordered. This disorder is strongly reduced if the growth is performed in a mixture of $\mathrm{N_2O}$ and $\mathrm{O_2}$ gas (growth type-A). Electric transport measurements confirmed that the first UC of LSCO15 next to the SLAO substrate is highly resistive and shows no sign of superconductivity for growth type-B, whereas it is superconducting for growth type-A. Furthermore, we found, rather surprisingly, that the conductivity of the LSCO15 UC next to the LCO capping layer strongly depends on the thickness of the latter. A LCO capping layer with 7~UCs leads to a strong localization of the charge carriers in the adjacent LSCO15 UC and suppresses superconductivity. The magneto-transport data suggest a similarity with the case of weakly hole doped LSCO single crystals that are in a so-called {\lq{cluster-spin-glass state}\rq}. We discussed several mechanisms that could lead to such a localization of holes that are embedded in a short-range ordered antiferromagnetic background for the case of a thick LCO capping layer with M=7 but not for a thin one with M=2.
\end{abstract}
\maketitle

\section{Introduction}
High-$\mathrm{T_c}$ superconductivity in the cuprates was first discovered in 1986 in the La-Ba-Cu-O compound~\cite{Bednorz1986}. Among various cuprates, the relatively simple crystal structure of $\mathrm{La_{2-x}Sr_xCuO_4}$ (La-214 system), the possibility to vary the Sr concentration, x, and thus the hole doping, p=x (for a perfect oxygen stoichiometry), over the entire superconducting part of the phase diagram and the availability of large single crystal have made it indispensable for many experimental studies. The undoped parent compound at x=0 is a charge-transfer insulator and a long-range ordered antiferromagnet (AF) with a N\'eel temperature of $\mathrm{T^N}{\approx}$325\,K~\cite{Kastner1998,Plakida2010}. The latter is rapidly reduced upon hole doping and vanishes at x$\geq$0.02. Nevertheless, a short range ordered AF kind of {\lq{cluster-spin-glass state}\rq} persists up to a doping level of x${\approx}$0.1, i.e., even into the superconducting part of the phase diagram~\cite{Cho1992,Niedermayer1998}. The superconducting critical temperature, $\mathrm{T_c}$, increases with doping in the underdoped regime to a maximal value of Tc${\approx}$38\,K at the optimal doping level of x=0.15-0.16, before it decreases again at higher doping in the so-called overdoped regime~\cite{Julien2003}. The normal state electronic properties in the underdoped regime are very unusual and seem to be governed by fluctuating and/or short-ranged magnetic and electronic orders that are presently not well understood and the subject of ongoing research. In the overdoped regime, especially at x$\geq$0.19, the normal state response becomes more conventional and Fermi-liquid-like~\cite{Kastner1998}. {\par}
The cuprates have a layered structure with quasi-two-dimensional $\mathrm{CuO_2}$ planes that host the superconducting charge carriers. They are separated by different kinds of insulating spacer layers, like the (La,Sr)O layers in La-214, which mainly serve as charge-reservoir. The superconducting and electronic properties accordingly exhibit a large anisotropy between the in-plane ($ab$-plane) and the out-of-plane ($c$-axis) directions. This allows one to realize superconductivity in a single $\mathrm{CuO_2}$ layer~\cite{Logvenov2009}, corresponding to half a unit cell of {\oplsco} (in the notation of the high temperature tetragonal phase with I4/mmm space group and lattice parameters of $a$=$b$=3.778 and $c$=13.237\,{\AA}~\cite{Francois1987}). This requires the growth of high quality ultra-thin films since, especially for $\mathrm{La_{2-x}Sr_{x}CuO_4}$, superconductivity depends critically on several factors, like the lattice mismatch between substrate and film, the oxygen stoichiometry, the intermixing of cations and other kinds of disorder and defects, as well as on a possible charge transfer across the interface (due to polar effects or a difference in the work functions of film and substrate, see supplementary information of Ref.~\cite{Gozar2008}). The ultrathin $\mathrm{La_{2-x}Sr_xCuO_4}$ (LSCO) films with the highest structural quality have been grown with the electron-beam and ozone or atomic-oxygen assisted co-evaporation~\cite{Sato1997} or molecular beam epitaxy (MBE)~\cite{Ruefenacht2003} techniques. Nevertheless, a superconducting single or double $\mathrm{CuO_2}$ layer has so far only be achieved by introducing an additional buffer layer between the substrate and the ultra-thin superconducting LSCO layer. This buffer layer is typically composed of several unit cells of undoped {\lco}~\cite{Bollinger2011} or strongly overdoped $\mathrm{La_{1.56}Sr_{0.44}CuO_4}$~\cite{Ruefenacht2003} that accommodate the detrimental effects as mentioned above. At the interface of such MBE-grown {\oplsco} (LSCO15) layers with a {\slao} (SLAO) substrate there is indeed an anomalous layer stacking of $\mathrm{AlO_2}$-(La,Sr)O-(La,Sr)O-(La,Sr)O-(La,Sr)O-$\mathrm{CuO_2}$ and the first 5 to 6 $\mathrm{CuO_2}$ layers next to the substrate are reported to be heavily distorted and contain significant amounts of oxygen vacancies~\cite{Zheng2014}. It has therefore been assumed that this buffer layer is not superconducting or even metallic and merely serves as a template for growing an ultrathin and superconducting LSCO15 layer on top. Nevertheless, it adds to the complexity of these artificial materials and thus introduces some uncertainty to the data interpretation.   {\par}
There have been only few attempts to grow LSCO15 films with the pulsed laser deposition (PLD)~\cite{Trofimov1994,Si1999,Rakshit2004}. To the best of our knowledge, there are no previous reports of PLD-grown ultra-thin LSCO15 films that are superconducting. Recent progress in this direction has been made by performing the PLD growth of LSCO15 films in a gas environment of $\mathrm{N_2O}$ instead of molecular oxygen~\cite{Das2014}. $\mathrm{N_2O}$ is a thermodynamically stable gas that provides a substantial amount of reactive atomic oxygen after collisions with the high-energetic plasma particles that are created during the ablation process~\cite{GUPTA1993}. Using this technique, a 10\,nm thick LSCO15 film with a superconducting transition in the R-T curve with $\mathrm{T_{c, on}}$ (onset) =40\,K and $\mathrm{Tc}$(R=0) =25\,K has been obtained. Notably, the stacking sequence at the substrate/film interface of such a film was found to be $\mathrm{AlO_2}$-(La,Sr)O-$\mathrm{CuO_2}$ (with a direct Al-O$_{\rm{apical}}$-Cu bond) and thus different from the one reported for the MBE-grown films in Ref.~\cite{Zheng2014}. Nevertheless, based on the \textit{in situ} growth monitoring with reflection high energy electron diffraction (RHEED), it was still found that at least the very first LSCO15 monolayer does not grow in a pure Frank-van der Merwe or \textit{layer-by-layer} mode and thus is likely to contain a significant amount of defects and dislocations that are detrimental to superconductivity. However, the growth of truly ultra-thin LSCO15 monolayers and the study of their superconducting properties have not been attempted in this previous work~\cite{Das2014}. {\par}
In this paper, we address this question and explore the superconducting properties of ultra-thin LSCO15 layers that are directly grown with PLD (without any buffer layer) on top of SLAO substrates. Using an atmosphere of pure $\mathrm{N_2O}$ gas as in Ref.~\cite{Das2014} we obtain SC LSCO15 layers down to a thickness of 2\,UCs. We also show that the PLD-growth can be further improved by using a mixture of $\mathrm{N_2O}$ and $\mathrm{O_2}$ gas such that even the first unit cell of LSCO15 on SLAO becomes superconducting. Finally, we show that the superconducting properties of such ultra-thin LSCO15 layers can be strongly affected by an additional LCO layer that is grown on top. In particular, we show that this LCO layer, which initially was intended to be a neutral capping layer, gives rise to a strong localization of the charge carriers in the neighboring LSCO15 layer if it is 7\,UCs thick whereas it does not have a noticeable effect if it consists of only 2\,UCs. We discuss several potential mechanisms for this surprising interface-effect.
\section{Experiment}
Series of bilayers (BLs) with N unit cells (UCs) of {\oplsco} and M\,UCs of {\lco} ([LSCO15\_N/LCO\_M]) have been grown with pulsed laser deposition (PLD). For comparison we have also prepared BLs for which the top layer consists of M\,UCs of $\mathrm{La_{1.94}Sr_{0.06}CuO_4}$ ([LSCO15\_N/LSCO6\_M]) or 20\,UCs of {\lcmo} ([LSCO15\_N/LCMO\_20]). The growth of these BLs was monitored with \textit{in situ} reflection high energy electron diffraction (RHEED) using a collimated 30\,keV electron beam at grazing incidence along the (1\,0\,0) or (0\,1\,0) crystallographic direction. The details of the PLD setup with \textit{in situ} RHEED can be found in Refs.~\cite{Malik2011,Das2014}. All depositions were made on commercially available (from MTI Corporation~\cite{MTI} or Crystal Gmbh~\cite{CRYSTAL}) {\slao} (SLAO) substrates with a (0\,0\,1) surface cut and a size of $5{\times}5{\times}0.5$\,$\rm{mm^3}$. {\par}
We used two different growth procedures which in the following are denoted as type-A and type-B. In both cases the substrate was preheated prior to the deposition for 1\,h at 730\,$\rm{^{\circ}C}$ in 0.11\,mbar of $\rm{N_2O}$ to cure the mechanically polished surfaces. {\par}
The deposition in A-type mode was carried out in a mixture of pure $\rm{N_2O}$ and $\rm{O_2}$ gas with partial pressures of 0.11 and 0.03\,mbar, respectively. The substrate temperature was kept at 730\,$\rm{^{\circ}C}$, and the target to substrate distance was set to about 5\,cm. For the LSCO15, LSCO6 and LCO layers the laser fluence was set to 1.0\,J/cm$\rm{^2}$ with a repetition rate of 2\,Hz. After each deposition, the gas mixture was flushed out from the PLD chamber with a flow of $\rm{O_2}$ gas for 25\,min at an equilibrium pressure of 0.3\,mbar. The $\rm{O_2}$ pressure was then increased to 1\,bar and the sample was subsequently cooled down to 550\,$\rm{^{\circ}C}$ at a rate of 5\,$\rm{^{\circ}C}$. At 550\,$\rm{^{\circ}C}$, the sample was annealed for 1\,h before it was rapidly cooled to room temperature. The BLs grown in this mode are labeled as [LSCO15\_N/LCO\_M]$\rm{^A}$. {\par}
For the B-type growth mode we used only pure $\rm{N_2O}$ gas with a partial pressure of 0.11\,mbar. The other growth parameters were the same as for the A-type. For the deposition of LCMO the laser fluence was set to 1.5\,J/cm$\rm{^2}$. The BLs of this type are marked as [LSCO15\_N/LCO\_M]$\rm{^B}$. {\par}
The x-ray diffraction (XRD) measurements were carried out with a four-circle diffractometer (Rigaku SmartLab) which is equipped with a 9\,kW rotating anode Cu-K${\alpha}$ source and a two-bounce Ge (2\,2\,0) monochromator. The wavelength resolution was ${\Delta}{\lambda}/{\lambda}$ = $3.8{\times}10^{-4}$. With symmetric $2{\theta}$-${\omega}$  scans we confirmed the epitaxial growth of the BLs and obtained the out-of-plane lattice parameters. The in-plane lattice parameters (and thus the strain condition of the layers) were determined by measuring the reciprocal space map (RSM) around the (1\,0\,11) Bragg peak of SLAO. {\par}
The cross-sectional high-resolution scanning transmission electron microscopy (STEM) images of a [LSCO15\_2/LCO\_7]$\rm{^A}$ BL were acquired in an aberration-corrected JEOL JEM-ARM200 CF, operated at 200\,kV, and equipped with a Gatan Quantum electron energy-loss spectrometer. All images presented here were obtained using a high-angle annular dark field (HAADF) detector. The specimens were prepared by conventional methods of grinding and Ar-ion milling. A thin layer of gold was evaporated on the STEM samples in order to prevent charging under the electron beam. {\par}
The resistance (R) measurements as a function of temperature and magnetic field were performed with a Physical Property Measurement System (PPMS from Quantum Design). A conventional four-probe method was used with the four wires glued with silver paint on the sample surface along a straight line. A dc current of 10\,$\mu$A was applied between the outer contacts, while the voltage was recorded between the two inner ones. The temperature and magnetic field were varied at rates of 2\,K/min and 0.01\,T/sec, respectively. For the measurement of the highly resistive samples we used an external Keithley Source Meter (Model: KE2602). The sheet resistances ({\rs}s) were calculated using the formula {\rs}=$\frac{\pi}{ln2}$R with the assumptions that the conduction mechanism is purely 2-dimensional and the samples are isotropic and homogeneous with respect to the charge conduction~\cite{Miccoli2015}. {\par}
The far-infrared (100-700\,$\rm{cm^{-1}}$) optical response of a [LSCO15\_2/LCO\_2]$\rm{^A}$ bilayer was determined using a home-built infrared ellipsometer~\cite{Bernhard2004}, based on a Bruker IFS-113 Fourier-Transform (FTIR) spectrometer, with a Hg discharge lamp as light source and a He-cooled bolometer as detector. The ellipsometer operates in the rotating analyzer mode with an optional static compensator that is based on a single internal reflection in a silicon prism. The spectra were taken at an angle of incidence, $\phi$= 75\,$^{\rm{\circ}}$. {\par}
The terahertz (3-70\,$\rm{cm^{-1}}$, 0.1-2.1\,THz) response of the same [LSCO15\_2/LCO\_2]$\rm{^A}$ BL was measured using a home-built time-domain THz (TD-THz) ellipsometer powered by a 100\,fs pulsed Ti:Sapphire laser by Menlo Systems, with photoconductive antennas as emitter and detector~\cite{Marsik2016}. The experiments were performed with different angles of incidence of 65, 70 and 75\,$^{\circ}$. {\par}
Both the FIR and the THz ellipsometers are equipped with CryoVac He-flow cryostats.
\section{Results and discussions}
\subsection{Structural characterization with \textit{in situ} RHEED}
\begin{figure}[!htbp] % Figure1: RHEED of type A
\centering
\includegraphics{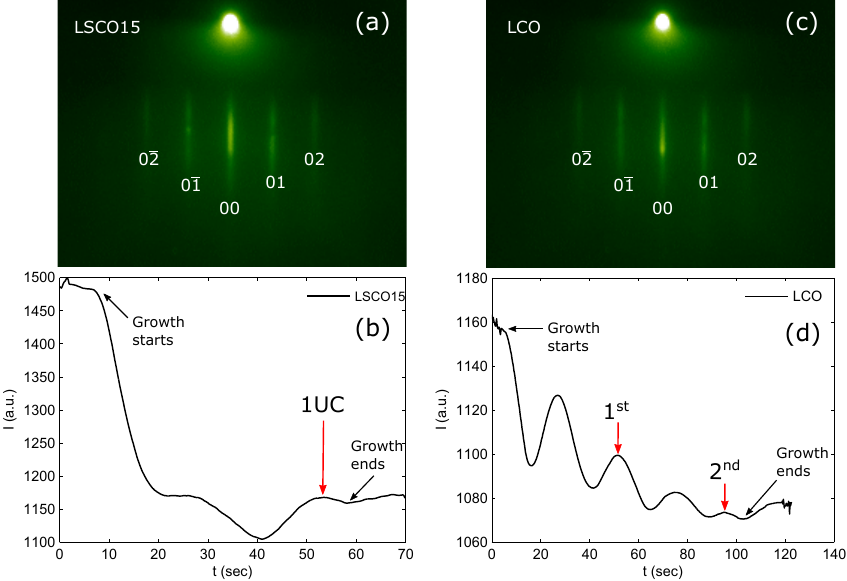}
\caption{\label{fig-rheedA} (a) RHEED pattern and (b) time evolution of the average intensity of the (0\,0) Bragg streak during the deposition of a 1\,UC thick LSCO15 layer on top of a SLAO substrate. The corresponding diagrams for a 2\,UCs thick LCO capping layer are shown in (c) and (d), respectively. The layers were grown with the growth type-A.}
\end{figure}

{\label{rheed}}
The structural quality and the growth mode of the LSCO15 and LCO layers have been monitored with \textit{in situ} RHEED. Representative RHEED patterns and the time evolution of the intensity of the (0\,0) Bragg peak during the growth of the A-type LSCO15 and LCO layers are shown in Fig.~\ref{fig-rheedA} for a [LSCO15\_1/LCO\_2]$\mathrm{^A}$ bilayer. The RHEED pattern in Fig.~\ref{fig-rheedA}(a) has been taken right after the growth of the LSCO15\_1 layer. It contains a set of well-defined Bragg peaks that are slightly elongated along the vertical direction. This typical streak-like appearance originates from defects of the film surface due to steps and terraces or other defects that enhance the roughness. The time-dependence of the intensity of the (0\,0) Brag peak during the growth is displayed in Fig.~\ref{fig-rheedA}(b). There are clear intensity oscillations that are characteristic of a \textit{layer-by-layer} or {\lq{Frank-van der Merwe growth mode}\rq}~\cite{Opel2012}. These oscillations arise from the periodic variation of the surface roughness as the add-atoms from the plasma-plume accumulate on the hot template and form a new monolayer of LSCO15. The RHEED intensity is at its maximum when the surface roughness is minimal, an oscillation thus corresponds to the growth of two ($\mathrm{La_{1-x} Sr_x}$ )O and one $\mathrm{CuO_2}$ atomic planes or half crystallographic UC. The two intensity maxima in Fig.~\ref{fig-rheedA}(b) thus confirm that the LSCO15 layer has a total thickness of 1\,UC. For the following LCO\_2 layer we obtained a similar RHEED pattern (Fig.~\ref{fig-rheedA}(c)) and the (0\,0) Bragg peak exhibits four intensity oscillations (Fig.~\ref{fig-rheedA}(d)) in agreement with the 2\,UCs layer thickness. {\par}
\begin{figure}[!htbp] % Figure2: RHEED of type B
\centering
\includegraphics{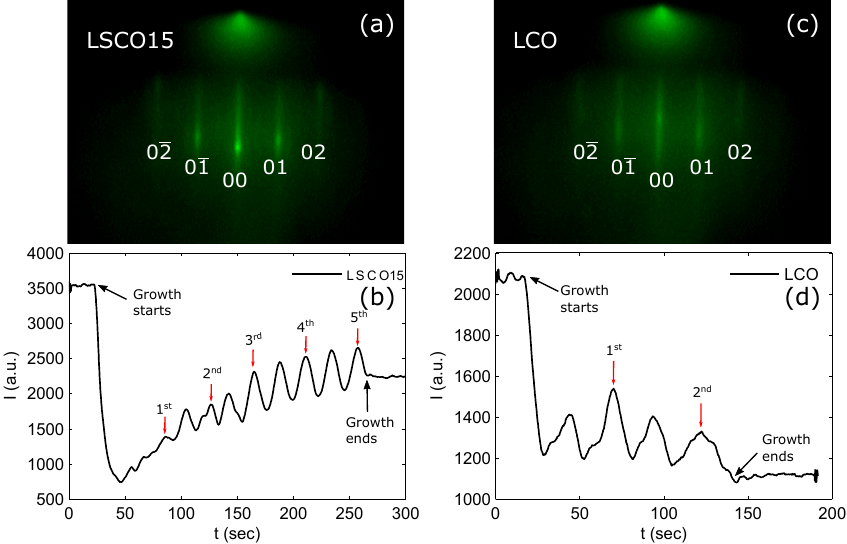}
\caption{\label{fig-rheedB} (a) RHEED pattern and (b) time evolution of the average intensity of the (0\,0) Bragg streak during the deposition of a 5\,UCs thick LSCO15 layer on top of a SLAO substrate. The corresponding diagrams for a 2\,UCs thick LCO capping layer are shown in (c) and (d), respectively. The layers were grown with growth type-B.}
\end{figure}
Figures.~\ref{fig-rheedB}(a)-(d) show the corresponding RHEED patterns and the evolution of the intensity of the (0\,0) Bragg peak during the growth of a [LSCO15\_5 /LCO\_2]$\mathrm{^B}$ bilayer. The most notable difference with respect to the [LSCO15\_1/LCO\_2]$\mathrm{^A}$ bilayer in Fig.~\ref{fig-rheedA} is that there are no clear intensity oscillations during the growth of the very first UC of LSCO15. For the LSCO15\_5 layer one expects ten intensity oscillations, whereas in Fig.~\ref{fig-rheedB}(b), only eight of them are clearly resolved. After a steep initial decrease of the RHEED intensity there are no clear oscillations in the range for which the growth of the first and the second monolayer occurs. The first growth oscillation is completely overdamped and there is only a faint hint of the second one. Clear growth oscillations are however seen for the third and the following monolayers. This suggests that the growth of the first two LSCO15 monolayers on top of the SLAO substrate does not follow a pure \textit{layer-by-layer} mode but rather a so-called {\lq{Stranski-Krastanov mode}\rq} which involves a mixture of a \textit{layer-by-layer} and 3D island growths~\cite{Opel2012}. The first LSCO15 monolayer (and partially even the second) thus is likely to contain a significant amount of misfit dislocations and/or stacking faults. Notably, a \textit{layer-by-layer} growth mode is recovered after the deposition of the first UC of LSCO15, as is evident from the relatively sharp intensity oscillations that are observed for the following LSCO15 and LCO layers. {\par}
Apparently, the gas environment has a strong influence on the growth mode of the first and even the second LSCO15 monolayer on top of the SLAO substrate. Whereas for the A-type growth condition a \textit{layer-by-layer} growth mode sets in already for the first LSCO15 monolayer, for the B-type it takes until the second or even third LSCO15 monolayer. Accordingly, for the B-type bilayers it can be expected that the first LSCO15 monolayer, and possibly even the second one, is strongly disordered and thus may not be superconducting or even metallic. As to the reason for such a large difference between the type-A and B conditions we can only speculate. For the condition B with pure $\rm{N_2O}$ gas it is known that the collisions of the high energetic particles from the plasma with the $\rm{N_2O}$ molecules lead to formation of atomic oxygen. For condition A with a mixture of $\rm{N_2O}$ and $\rm{O_2}$ there may be an additional interaction of this atomic oxygen with the $\rm{O_2}$ molecules which results in ozone ($\rm{O_3}$) that is even more reactive and well known to aid the growth of high quality LSCO~\cite{Bozovic2002}.
\begin{figure}[!htbp] % Figure3: XRD of type A
\centering
\includegraphics{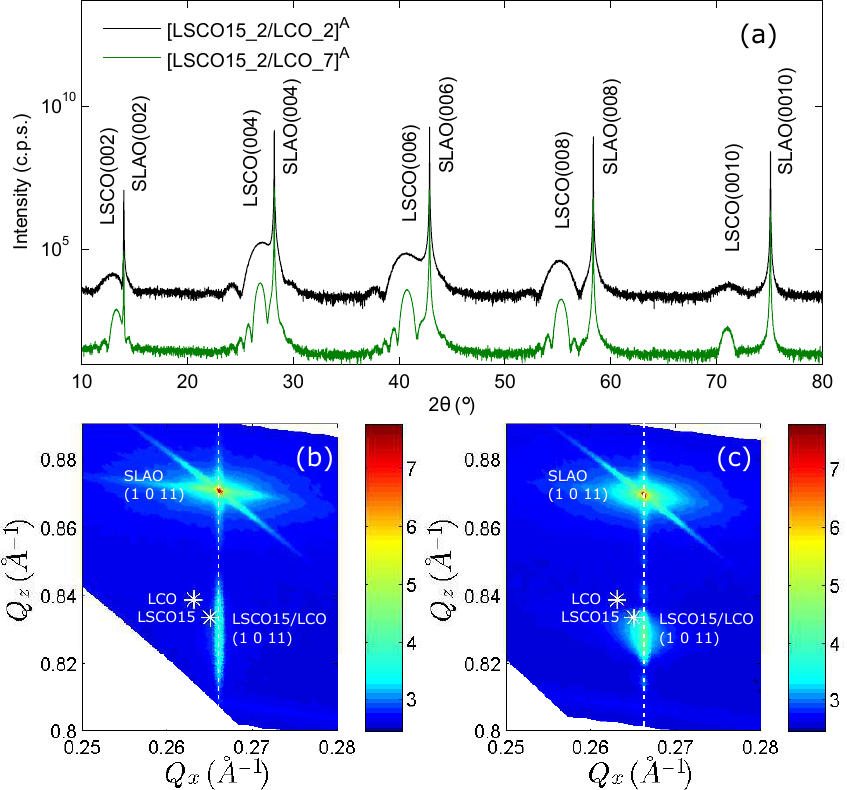}
\caption{\label{fig-xrd} (a) Symmetric 2$\theta$-$\omega$ x-ray diffraction patterns along the [0\,0\,L]-direction of the [LSCO15\_2/LCO\_2]$\mathrm{^A}$ and [LSCO15\_2/LCO\_7]$\rm{^A}$ bilayers. (b) and (c) Corresponding reciprocal space maps (RSMs) around the (1\,0\,11) SLAO peak for [LSCO15\_2/LCO\_2]$\rm{^A}$ and [LSCO15\_2/LCO\_7]$\rm{^A}$, respectively. The dashed white line connects the (1\,0\,L)-peaks of SLAO). The asterisks show the location of the corresponding Bragg peaks for bulk LSCO15 and LCO. }
\end{figure}
\subsection{Structural characterization with XRD}{\label{xrd}}
Figure~\ref{fig-xrd}(a) shows the XRD patterns for symmetric 2$\theta$-$\omega$ scans of the samples [LSCO15\_2/LCO\_2]$\rm{^A}$ and [LSCO15\_2/LCO\_7]$\rm{^A}$ on (0\,0\,1)-oriented SLAO substrates. Only the (0\,0\,L) Bragg peaks can be observed which confirms the epitaxial growth of the samples with the $c$-axis perpendicular to the surface. Since LSCO15 and LCO have very similar lattice constants, their Bragg peaks that are broadened due to the finite thickness of the layers cannot be distinguished within our experimental resolution of ${\Delta}{\theta}{\approx}0.01$$\rm{^{\circ}}$. For the thicker [LSCO15\_2/LCO\_7]$\rm{^A}$ bilayer there are clear intensity oscillations around the intense Bragg peaks that are linked to the total thickness of the bilayers and testify for the sharpness of the interfaces and the low surface roughness. For the [LSCO15\_2/LCO\_2]$\rm{^A}$ bilayer these oscillations are not resolved due to the intrinsic broadening of the Bragg-peaks that arises from the smaller total layer thickness. The average $c$-axis lattice parameter, $c_{\rm{expt}}$, as calculated from the (0\,0\,8)-peaks, amounts to 13.30(2) and 13.27(1)\,{\AA}  for [LSCO15\_2/LCO\_2]$\rm{^A}$ and [LSCO15\_2/LCO\_7]$\rm{^A}$, respectively. These are somewhat larger than in bulk LSCO15 ($c_{\rm{bulk}}$ =13.24\,{\AA}) or LCO ($c_{\rm{bulk}}$ =13.12\,{\AA}) as expected since the SLAO substrate gives rise to a compressive strain along the in-plane directions ($a$-axis, $b$-axis). {\par}
Figures~\ref{fig-xrd}(b) and~\ref{fig-xrd}(c) show corresponding reciprocal space maps (RSMs) around the SLAO (1\,0\,11) Bragg peak. For the [LSCO15\_2/LCO\_2]$\rm{^A}$ bilayer the corresponding Bragg peak of the LSCO15/LCO stack is nearly symmetric with respect to the $h$=$1$ line that is defined by the SLAO (1\,0\,L)-peaks. This confirms that the LSCO15 and the LCO layer are both fully strained, i.e., there is hardly any strain relaxation. For the [LSCO15\_2/LCO\_7]$\rm{^A}$ bilayer there is a weak asymmetry towards lower-$Q_x$ and higher-$Q_z$  values that indicates a partial strain relaxation. Nevertheless, for both bilayers the average in-plane lattice parameter ($a_{\rm{expt}}$) of the LSCO15/LCO stack amounts to 3.76(1)\,{\AA}.
\subsection{Structural characterization with STEM}
\begin{figure}[!htbp] % Figure4: STEM of type A
\centering
\includegraphics{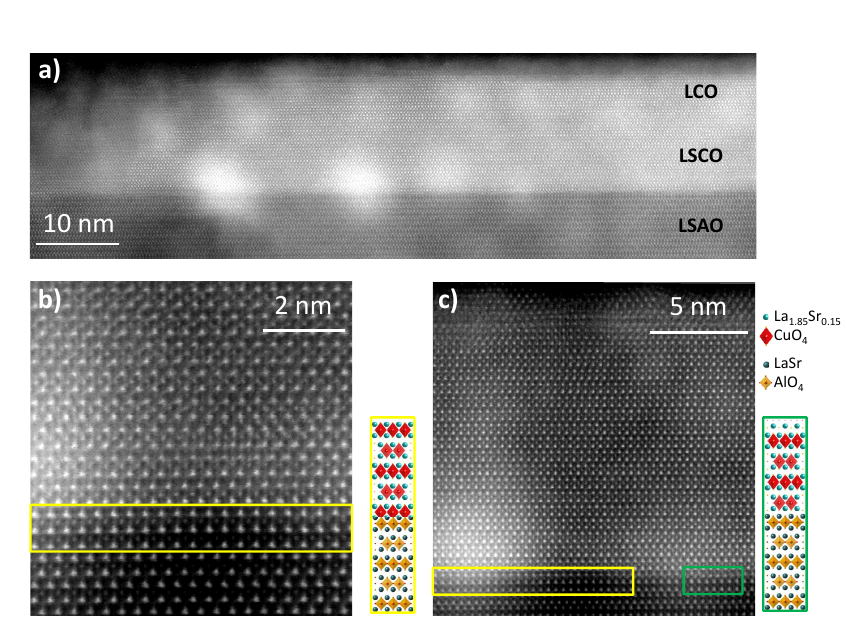}
\caption{\label{fig-stem} (a) Low magnification HAADF image of a [LSCO15\_2/LCO\_7]$\rm{^A}$ bilayer grown on a SLAO substrate. (b) High magnification HAADF image of the LSCO15/SLAO interface. The interface area is highlighted by a yellow rectangle. The proposed structural model for the interface is indicated on the right (yellow rectangle). (c) HAADF image showing a larger lateral scale view of a part of the interface which exhibits an area with a different termination, as indicated in the model on the right (green rectangle). The patches of bright contrast are associated with nanometric scale Au droplets that have been evaporated on the sample surface to prevent charging effects due to the electron beam. }
\end{figure}

{\label{stem}}
Figure~\ref{fig-stem} displays representative HAADF images of a [LSCO15\_2/LCO\_7]$\rm{^A}$ bilayer grown on a SLAO substrate. The low magnification images in Fig.~\ref{fig-stem}(a) show flat and uniform layers with no major defects or secondary phases. The LSCO15 and LCO layers cannot be distinguished from each other due to the similarity of their atomic structure and their average composition. The bright spots which can be observed on the images are associated with the discontinuous nature of an ultrathin Au layer that was evaporated on the surface of the bilayer prior to the sample preparation in order to prevent charging effects of the highly insulating sample (substrate SLAO) under the electron beam. Fig.~\ref{fig-stem}(b) shows a high magnification view of the LSCO15/SLAO interface. Samples are epitaxial and interfaces are free of major defects. The characteristic structure of three atomic planes consisting of $\mathrm{La_{1.85}Sr_{0.15}}$ and LaSr on the LSCO15 and SLAO side, respectively, has been highlighted with a yellow rectangle. This type of interface structure provides the contact via the apical oxygen between the $\mathrm{CuO_4}$ and $\mathrm{AlO_4}$ octahedra – see the structural model on the right (inset in yellow rectangle). In spite of this epitaxy and the good matching of LSCO15 and SLAO lattice constants ($a\mathrm{_{LSCO15}^{bulk}}$=3.778\,{\AA}; $a\mathrm{_{SLAO}^{bulk}}$=3.756\,{\AA}), occasional discontinuities probably related to mismatch are observed, as shown in Fig.~\ref{fig-stem}(c). On occasion, a different interface atomic sequence is present (highlighted by a green rectangle) where two rocksalt-like atomic La/Sr planes impede the contact between $\mathrm{CuO_4}$ and $\mathrm{AlO_4}$ octahedra, see the adjacent model in the green rectangle.  {\par}
Very similar HAADF images have been obtained for a [LSCO15\_5/LCO\_2]$\mathrm{^B}$ bilayer (not shown here). They show the same type of interface termination between the SLAO substrate and the LSCO15 layer with a majority of regions with a direct contact between the $\mathrm{CuO_4}$ and $\mathrm{AlO_4}$ octahedra via the apical oxygen and a minority of regions with two rocksalt-like atomic La/Sr planes at the interface. A meaningful quantitative analysis of the distribution of these regions with different interface termination was unfortunately not possible since only a very small part of the sample could be studied with the TEM technique. This is related to difficulties with the sample preparation that are linked to the poor mechanical properties of the SLAO substrates (which are very brittle).
\subsection{Superconductivity in ultrathin LSCO15 layers of type-A}
\begin{figure}[!htbp] % Figure6: Ellipsometry
\centering
\includegraphics{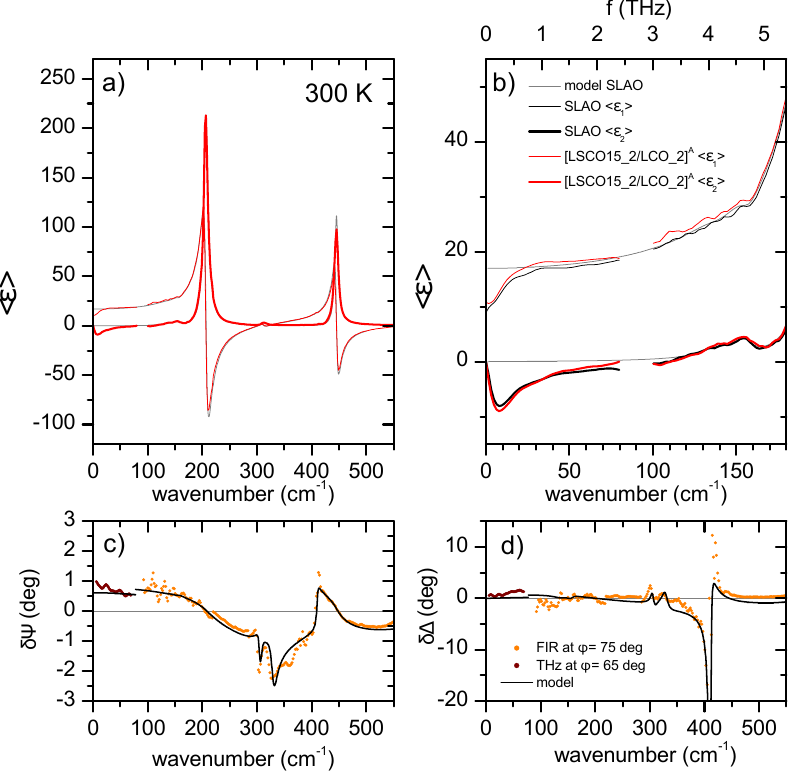}
\caption{\label{fig-six} (a) Room-temperature spectra of the real and imaginary parts of the pseudo-dielectric function, $\ev{{\varepsilon}_1}$ (red thin line) and $\ev{{\varepsilon}_2}$ (red thick line), of a [LSCO15\_2/LCO\_2]$\rm{^A}$ bilayer on a SLAO substrate measured by TD-THz ellipsometry (up to 80\,cm$^{-1}$) at $\phi$=\,65$^\circ$ and FTIR ellipsometry (above 100\,cm$^{-1}$) at ${\phi}$=\,75$^\circ$. The spectra are governed by the strong phonon modes of the SLAO substrate and the much weaker response of the bilayer is hardly visible. (b) Magnification of the THz regime for which a small difference between the spectra of the bilayer (red lines) and the bare substrate (black lines) can be resolved. The gray lines show the expected THz response of the bare SLAO substrate as obtained from the modelling of the FTIR ellipsometry data above 100\,cm$^{-1}$. The difference with respect to the measured spectra which increases towards low frequency originates from diffraction effects due to the finite sample size. (c) and (d) Difference spectra of the ellipsometric angles of the bilayer sample with respect to the bare SLAO substrate, ${\delta}{\Psi}$=${\Psi}$(bilayer/substrate)$-$${\Psi}$(substrate) and ${\delta}{\Delta}$=$\Delta$(bilayer/substrate)$-$$\Delta$(substrate), for which the contribution of the bilayer is more easily seen (solid symbols). The solid black lines show a model calculation of the differential spectra for a metallic thin film with $d$=2.6\,nm on a SLAO substrate (see the text for a discussion). }
\end{figure}

{\label{rt-typeA}}
\begin{figure}[!htbp] % Figure5: RT of type A
\centering
\includegraphics{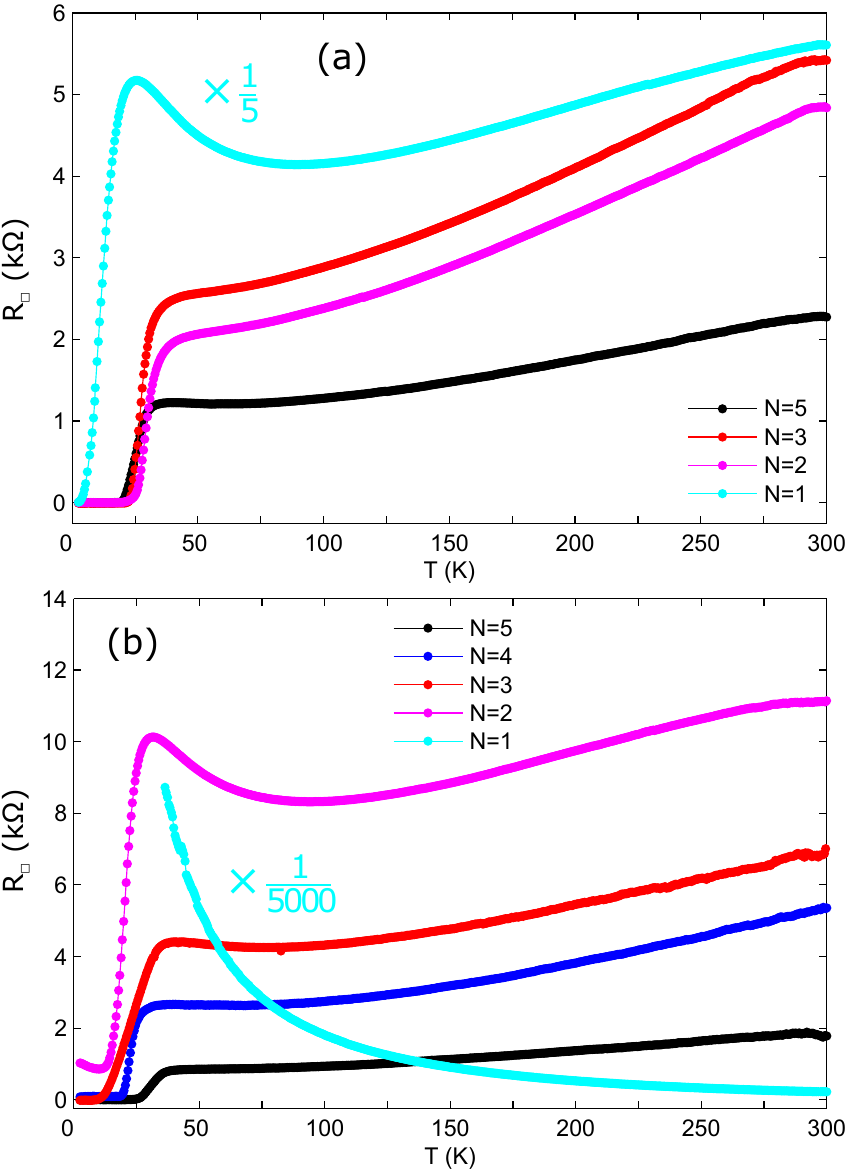}
\caption{\label{fig-rt-bl} (a) Sheet resistance versus temperature ({\rs}-T) curves of a series of [LSCO15\_N / LCO\_2]$\rm{^A}$ bilayers. (b) Corresponding {\rs}-T curves of [LSCO15\_N/LCO\_2]$\rm{^B}$ bilayers.}
\end{figure}
Figure~\ref{fig-rt-bl}(a) displays the sheet resistance versus temperature ({\rs}-T) curves of the A-type bilayers [LSCO15\_N/LCO\_2]$\rm{^A}$ that were grown in the mixture of $\rm{N_2O}$ and $\rm{O_2}$ gas. Here the thickness of the LSCO15 layers is varied between 1 and 5\,UCs while the LCO layer always has a thickness of 2\,UCs. The insulating LCO layer on top has been added as a capping layer to protect the ultrathin LSCO15 layer underneath from degradation due to the direct interaction with the ambient atmosphere and with the solvent of the silver paint that was used to make the contacts for the \textit{ex situ} resistance measurements. {\par}
Except for the N=1 bilayer, the samples have a reasonably low resistance and a metallic temperature dependence. They also exhibit a superconducting (SC) transition with similar onset and zero resistance temperatures of  $\mathrm{T_{c, on}}$$\approx$35\,K and  $\mathrm{T_{c}}$({\rs}=0)$\approx$22\,K, respectively. For the N=1 bilayer there is an onset of a SC transition at  $\mathrm{T_{c, on}}$$\approx$20\,K but a zero resistance state is not reached down to the lowest measured temperature of 2\,K. In addition, in the normal state, the resistance is significantly higher than for the N=2-5 bilayers and exhibits an upturn below about 75\,K which provides evidence for a weak localization of the charge carriers. Together with the incomplete superconducting transition this suggests that the 1\,UC thick LSCO15 layer is not a homogeneous superconductor but rather consists of superconducting islands that are weakly coupled via the Josephson effect~\cite{Baturina2007,Adkins1980}. The zero resistance state below $\mathrm{T_{c}}$({\rs}=0)$\approx$22\,K of the bilayers with N$>$1 is certainly also not a firm proof for a macroscopically homogeneous superconducting state of the LSCO15 layer, since a similar {\rs}-T curve may even be obtained from a filamentary superconductor. {\par}
More direct evidence for a bulk-like superconducting response of the [LSCO15\_2/LCO\_2]$\rm{^A}$ bilayer has been obtained from an additional terahertz (THz) and infrared spectroscopy study. {\par}
Figure~\ref{fig-six}(a) shows the measured room temperature spectra of the real and imaginary parts of the pseudo-dielectric function, $\ev{{\varepsilon}_1}$ and $\ev{{\varepsilon}_2}$, in the THz- and FIR range for the [LSCO15\_2/LCO\_2]$\rm{^A}$ bilayer (red lines). The spectra are governed by two strong infrared-active phonon modes of the SLAO substrate around 207 and 445\,cm$^{-1}$~\cite{Humlicek2000} such that much weaker response of the [LSCO15\_2/LCO\_2]$\rm{^A}$ bilayer is hardly visible. A small yet clearly resolved difference between the spectra of the bilayer (red lines) and the bare SLAO substrate (black lines) is still visible in the magnified view of the THz regime that is shown in Fig.~\ref{fig-six}(b). The main signature is a small upward shift of $\ev{{\varepsilon}_1}$ that is consistent with a metallic conductivity of the bilayer~\cite{Marsik2016}. Also shown by the gray line is the expected THz response of the bare SLAO based on the modeling of the FIR data above 100\,cm$^{-1}$. The difference with respect to the measured spectra (black lines) arises mainly from diffraction-related artifacts and is further discussed below in the context of the temperature dependent response. {\par}
\begin{figure}[!htbp] % Figure7: Ellipsometry
\centering
\includegraphics{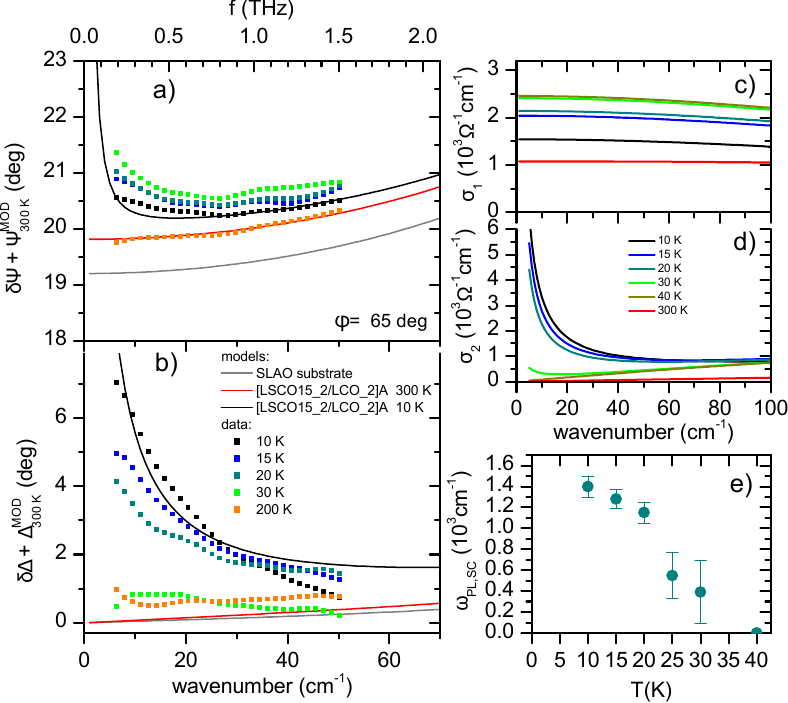}
\caption{\label{fig-seven} (a) and (b) Temperature dependence of the THz response of the [LSCO15\_2/LCO\_2]$\rm{^A}$ bilayer in terms of the spectra of the ellipsometric angles, $\Psi$ and $\Delta$. Solid symbols show the sum of the measured temperature difference spectra with respect to room temperature, ${\delta}{\Psi}$=$\Psi$(T)$-$$\Psi$(300\,K) and ${\delta}{\Delta}$=$\Delta$(T)$-$$\Delta$(300\,K), and the model spectra at room temperature, $\Psi\mathrm{_{300\,K}^{MOD}}$, $\Delta\mathrm{_{300\,K}^{MOD}}$ (red lines). This procedure is used to remove the almost temperature independent diffraction effects that are shown in Fig.~\ref{fig-six}(b) for the spectra at 300\,K. The solid lines show the best fits with the model of a 2.6\,nm thick metallic and superconducting layer (at T$<\mathrm{T_c}$) on a SLAO substrate for which the response is described with the Drude-model in eq.~\eqref{eq-one}. (c) and (d) Temperature dependence of the spectra of the real- and imaginary parts of the optical conductivity, $\sigma_1$ and $\sigma_2$, of the LSCO15\_2 layer as obtained from the best fits to the data in (a) and (b). (e) Temperature dependence of the plasma frequency of the superconducting charge carriers that is proportional to the square root of the superconducting condensate density, $\omega_\mathrm{PL,SC}\approx\sqrt{n_s}$.}
\end{figure}
Figures~\ref{fig-six}(c) and~\ref{fig-six}(d) reveal that the contribution of the bilayer is more readily seen in the difference spectra of the ellipsometric angles, ${\delta}{\Psi}$ and ${\delta}{\Delta}$, with respect to the bare substrate (solid symbols). The black lines furthermore show that a reasonable fit of these differential data can be obtained with a model which compares the response of a 2.6\,nm thick, metallic layer on a SLAO substrate with the one of a bare SLAO substrate. This metallic layer accounts for the LSCO15\_2 layer and is described by the Drude model of eq.~\eqref{eq-one}. The comparably much weaker response of the insulating LCO\_2 layer on top is neglected. From the best fit we obtain a plasma frequency of $\omega_{\rm{PL}}$=6600\,cm$^{-1}$ and a broadening parameter $\gamma$=680\,cm$^{-1}$ which suggest a dc value of $\sigma^{\mathrm{dc}}$$\approx$1100\,$\Omega^{-1}$cm$^{-1}$. The latter agrees reasonably with the dc transport data in Fig.~\ref{fig-rt-bl}(a) which yield an estimate of $\sigma^{\mathrm{dc}}$$\approx$781\,$\Omega^{-1}$cm$^{-1}$ at 300\,K. {\par}
This model of the room temperature spectra serves as a baseline for the following analysis of the temperature dependent response of the A-type LSCO15\_2 layer. As can be seen in Fig.~\ref{fig-six}(b) and was already mentioned above, the raw THz ellipsometry data below 50\,cm$^{-1}$ are increasingly affected by diffraction effects~\cite{Humlicek2004,Marsik2016}. To remove these artifacts, we show in Figs.~\ref{fig-seven}~(a),~(b) the temperature dependence of the ellipsometric angles $\Psi$ and $\Delta$ (solid symbols) in terms of the sum of the measured difference spectra with respect to room temperature, ${\delta}{\Psi}$=$\Psi$(T)$-$$\Psi$(300\,K) and ${\delta}{\Delta}$=$\Delta$(T)$-$$\Delta$(300\,K), and the model spectra at room temperature, ${\Psi}_{\rm{300\,K}}^{\rm{MOD}}$, ${\Delta}_{\rm{300\,K}}^{\rm{MOD}}$ (red lines in Figs.~\ref{fig-seven}~(a),(b)). We have previously shown that for an ultrathin film on a dielectric substrate, the changes of $\Psi$ are related to the real part of the optical conductivity of the film, $\sigma_1$, and the ones of $\Delta$ reflect the changes of the imaginary part, $\sigma_2$~\cite{Marsik2016}. {\par}
An estimate of the temperature dependence of the optical conductivity, $\sigma(\omega)$, of the thin film has been obtained by fitting the spectra in Figs.~\ref{fig-seven}(a)~and~(b) with a simple model containing two Drude terms:
\begin{equation}\label{eq-one}
\sigma(\omega)=i\sigma_0(
\frac{\omega^2_\mathrm{PL}}{\omega+i\gamma}
+
\frac{\omega^2_\mathrm{PL,SC}}{\omega}
),
\end{equation}
where the first term describes the response of the normal carriers with the plasma-frequency ${\omega}_{\rm{PL}}$ and broadening $\gamma$, and the second term the loss-free response of the superconducting condensate with a plasma-frequency ${\omega}_{\rm{PL,SC}}$ (that is finite only below $\rm{T_c}$). The obtained spectra of the real and the imaginary parts of the optical conductivity of the A-type LSCO15\_2 layer are displayed in Figs.~\ref{fig-seven}~(c), (d).  They are fairly consistent with the previously reported THz optical response of $\mathrm{La_{2-x}Sr_xCuO_4}$ thin films~\cite{Bilbro2011,Nakamura2012}. In the normal state, ${\omega}_{\rm{PL}}$ has been fixed to the room temperature value of 6600\,cm$^{-1}$. The only fit parameter has been the broadening, ${\gamma}$, which decreases from $\gamma$=680\,cm$^{-1}$ at 300\,K to ${\gamma}$=300\,cm$^{-1}$ at 40\,K. The corresponding low-frequency conductivity, ${\sigma}_1$, increases towards lower temperature and reaches a maximum value of about 2450\,${\Omega}^{-1}\rm{cm}^{-1}$ at 30-40\,K. The dc transport data in Fig.~\ref{fig-rt-bl} once more agree with such a more than two-fold increase of the dc conductivity between 300 and 40\,K. In the SC state below $\rm{T_c}$, the broadening parameter was fixed to the value at 40\,K ($\gamma$=300\,cm$^{-1}$) whereas the plasma-frequencies, $\omega_{\rm{PL}}$ and $\omega_{\rm{PL,SC}}$, were used as fit parameters. The evolution of $\omega_{\rm{PL,SC}}$, which is proportional to the square root of the superconducting condensate density, $n_s$, is displayed in Fig.~\ref{fig-seven}(e). It exhibits a fairly rapid increase below $\rm{T_c}$ as expected for a quantity that is proportional to the amplitude of the superconducting order parameter. The extrapolated low-temperature value amounts to $\omega_{\rm{PL,SC}}$(T${\rightarrow}0$)$\approx$1600\,cm$^{-1}$. It is much smaller than the value of $\omega_{\rm{PL,SC}}$(T${\rightarrow}0$)$\approx$6000\,cm$^{-1}$ that was reported from the FIR spectroscopy data of an almost optimally doped LSCO14 single crystal~\cite{Startseva1999}. This suggests that the SC condensate density of the ultrathin LSCO film (${\omega_{\rm{PL,SC}}}^2{\sim}n_s$) is about 14 times lower than in the single crystal. However, this is not too surprising for such an ultrathin layer of a cuprate superconductor with a d-wave symmetry of the order parameter for which any kind of scattering, for example due to the surface roughness, gives rise to destructive interference effects that lead to a rapid suppression of the condensate density~\cite{Bernhard1996}. Moreover, Kosterlitz-Thouless-type fluctuation effects due to the unbinding of vortex-antivortex pairs tend to suppress the superconducting condensate in such ultra-thin films~\cite{Nelson1977}. Based on the suppression of $\rm{T_c}$ (R=0) from 38\,K in the bulk to about 22\,K in this thin film and the almost linear relationship between $\rm{T_c}$ and $n_s$ in the cuprates~\cite{Emery1995} we estimate that this effect reduces the condensate density by about 40\%. The observation of a loss-free (inductive) response due to a sizeable superconducting condensate is therefore a strong indication that the superconducting state of this ultrathin film is fairly homogenous and certainly not of filamentary type.
\subsection{Superconductivtiy in ultrathin LSCO15 layers of type-B}{\label{rt-typeB}}
Figure~\ref{fig-rt-bl}(b) shows the {\rs}-T curves of the corresponding type-B bilayers [LSCO15\_N/LCO\_2]$^{\rm{B}}$ with N=1-5 that have been grown in a pure $\rm{N_2O}$ atmosphere. For this series only the bilayers with N$\geq$3 show a metallic behavior and a complete superconducting transition with zero resistance at T$>$2\,K. They have similar onset temperatures of $\rm{T}_{\rm{c,on}}$$\approx$35\,K,  though the value of $\rm{T_c}$({\rs}=0) increases with N from 13\,K to 26\,K. The N=2 bilayer exhibits a resistive upturn and an incomplete SC transition similar to the N=1 sample of the A-type series. Furthermore, the {\rs}-T curve of the N=1 bilayer is characteristic of an insulator-like behavior. It exhibits a large sheet resistance of about  1128\,k$\Omega$ at 300\,K that rises steeply towards low temperature and shows no sign of a SC transition. {\par}
These marked differences in the transport properties of the A- and B-type bilayers suggest that there exists an offset of about 1\,UC of LSCO15 in the minimum thickness that is needed to obtain a metallic behavior and a macroscopic superconducting state. This can be understood in terms of the different growth mode of the first UC of LSCO15 (or the first two LSCO15 monolayers) next to the SLAO substrate that has been identified with the \textit{in~situ} RHEED (see section~\ref{rheed}). A two-dimensional \textit{layer-by-layer} growth mode, which yields sharp interfaces and a high structural quality, starts here from the very first LSCO15 monolayer (half of a UC) for the type-A samples but only from the third monolayer for the type-B. It thus can be expected for the latter that the first UC of LSCO15 contains a large amount of disorder and defects which destroy superconductivity and even a metallic response. Even the second LSCO15 UC of the B-type bilayer and also the first one of the A-type bilayer seem to be somewhat disordered since their carriers are weakly localized at low temperature and the superconducting state seems to be only of a granular type. The following UCs of LSCO15 with N$\geq$2 for the type-A and N$\geq$3 for the type-B bilayers are all metallic and have a complete superconducting transition with onset temperatures of about 35\,K or even higher.
\subsection{Impact of a thicker LCO capping layer}\label{thick-lco}
\begin{figure*}[htb] % Figure8: RT of  BLs with thick LCO
\centering
\includegraphics{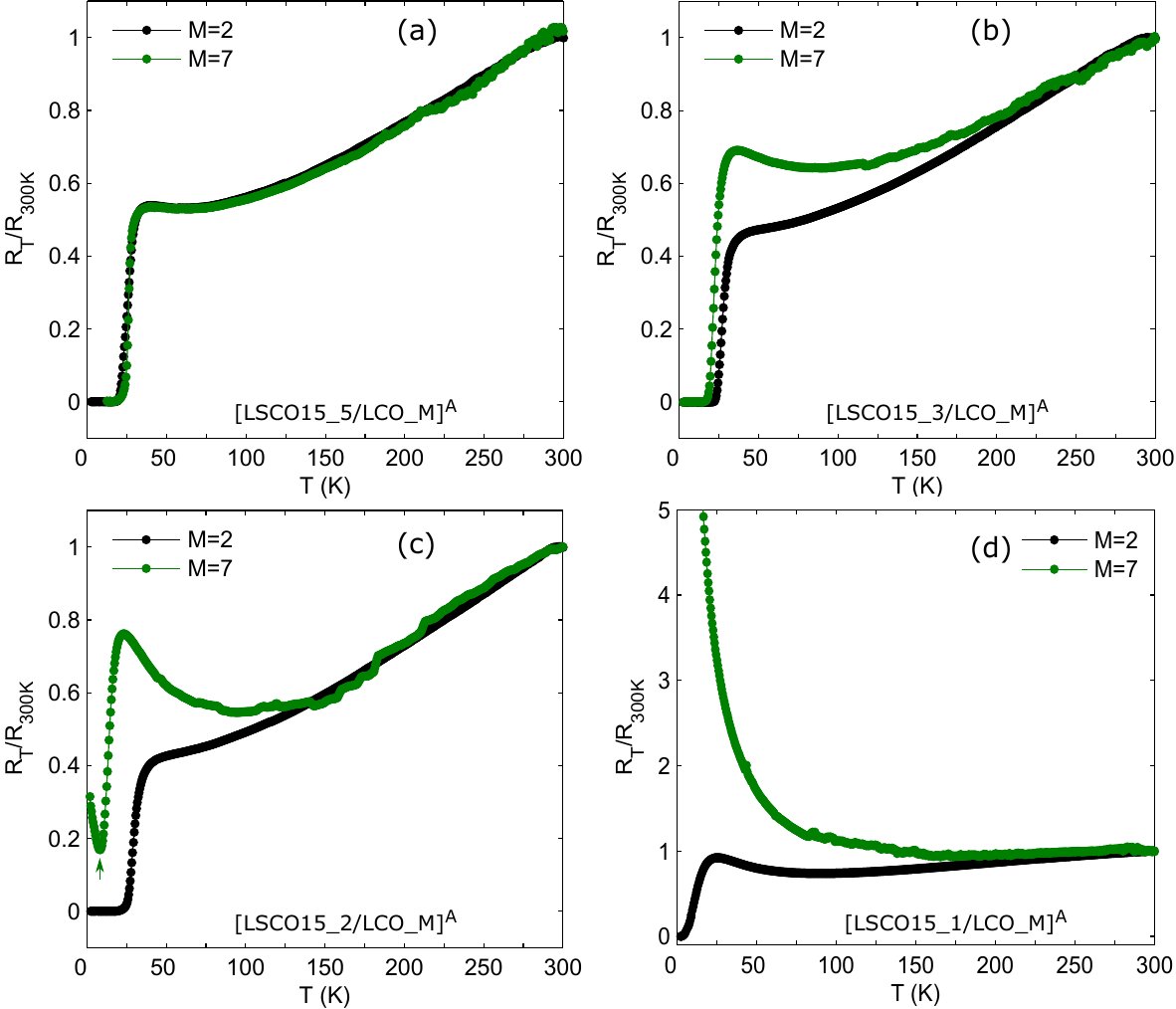}
\caption{\label{fig-rt-thickLCO} Temperature  dependence of the normalized resistance, ($\rm{R_T}$/$\rm{R_{300\,K}}$), of the bilayers (a) [LSCO15\_5/LCO\_M]$\rm{^A}$,  (b) [LSCO15\_3/LCO\_M]$\rm{^A}$, (c) [LSCO15\_2/LCO\_M]$\rm{^A}$, and (d) [LSCO15\_1/LCO\_M]$\rm{^A}$.}
\end{figure*}
To our surprise, we found that the metallic and superconducting response of the LSCO15 layers is also strongly dependent on the thickness of the LCO cap layer. This is shown in Figs.~\ref{fig-rt-thickLCO}(a)-(d) in terms of the R-T curves of two series of type-A bilayers [LSCO15\_N/LCO\_M]$\rm{^A}$ with N=1, 2, 3 and 5, and M=2 and 7. For a better comparison, the R-T curves are normalized with respect to the value at 300\,K. For N=5, the metallic normal state resistance and the SC transition do not depend on the thickness of the LCO capping layer. However, for the samples with N$\leq$3 there are clear difference between the samples with M=2 and 7 that become larger as the thickness of the LSCO15 layer is reduced. For N=3 and M=7 there is a weak resistive upturn in the normal state and a slightly reduced SC transition temperature as compared to the N=3 and M=2 sample for the normal state resistance keeps decreasing all the way to $\rm{T_c}$. For N=2 and M=7, the resistive upturn is even more pronounced than for N=3 and M=7. The superconducting transition is now strongly suppressed and it remains incomplete. The resistive upturn toward very low temperature (marked by an arrow in Fig.~\ref{fig-rt-thickLCO}(c)) is indicative of a granular superconducting state~\cite{Baturina2007,Frydman2002,Jaeger1989}. Finally, for the N=1 and M=7 sample, there is only an insulator-like upturn of the resistance with no sign of a superconducting transition. The R-T curve is in fact well described by a 2D variable-range-hopping (VRH) model which indicates that the charges carriers get localized at low temperature and the charge transport occurs only via hopping processes (see Appendix~\ref{app-vrh}). {\par}
\begin{figure}[!htbp] % Figure9: RH of  BL
\centering
\includegraphics{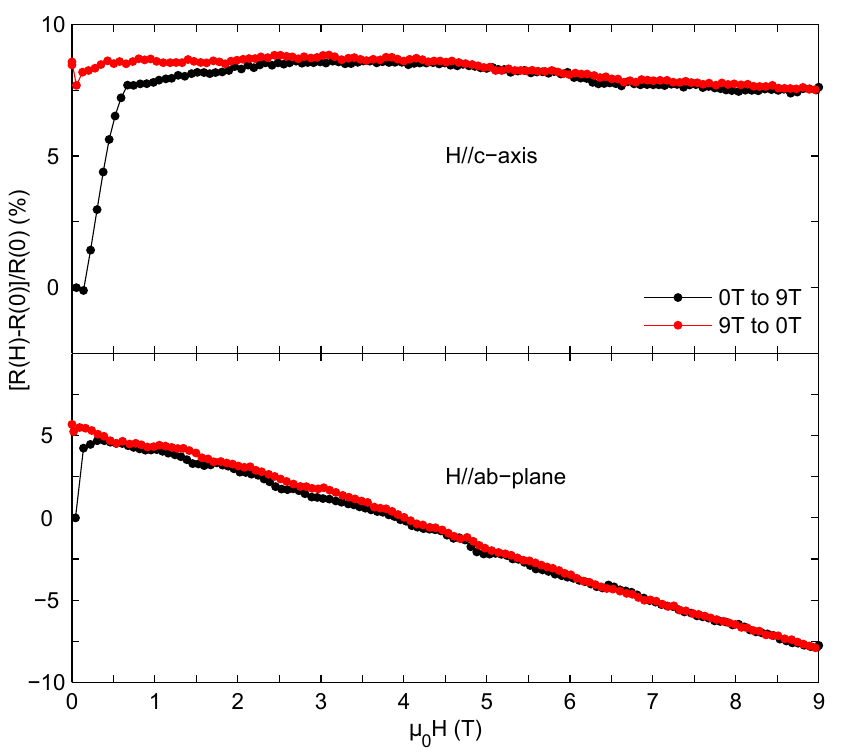}
\caption{\label{fig-rh} Magneto-resistance, [R(H)-R(0)]/R(0), of the [LSCO15\_1/LCO\_7]$\rm{^A}$ bilayer measured at 2\,K after zero field cooling. In the upper and lower panels the magnetic fields has been applied along the $c$-axis and the $ab$-plane of the sample, respectively.}
\end{figure}
Overall these data show that in addition to the first UC of LSCO15 next to the SLAO interface, for which the carriers are weakly localized and SC is of a granular type, in the first UC of LSCO15 next to the LCO capping layer with M=7 the charge carriers are strongly localized and superconductivity is fully suppressed. Notably, this additional interface effect occurs only when the thickness of the LCO capping layer exceeds a certain threshold, i.e., for M=7 but not for M=2. {\par}
We have further explored the nature of the charge carrier localization by measuring the magneto-resistance of the [LSCO15\_1/LCO\_7]$\rm{^A}$ bilayer. Figs.~\ref{fig-rh}(a)~and~(b) display the R-H curves at 2\,K after zero field-cooling with the magnetic field applied along the directions parallel (H${\parallel}ab$) and perpendicular (H${\parallel}c$) to the $\rm{CuO_2}$ planes, respectively. For both field orientations, there is an initial steep increase of R, corresponding to a positive magneto-resistance (pMR), that is followed by a more gradual decrease of R toward high magnetic field due to a negative magnetoresistance (nMR). Furthermore, upon decreasing the field again, there is a pronounced hysteresis effect in the low-field regime where R does not return to the initial value. {\par}
All the three effects, i.e., the pMR and the hysteresis at low field as well as the nMR at high field were previously observed in strongly underdoped LSCO single crystals that exhibit a so-called {\lq{cluster-spin-glass}\rq} state in which the holes reside at the boundaries of undoped (or very weakly doped) regions that host a short-range antiferromagnetic order~\cite{Raicevic2010}. The microscopic mechanism of these MR-effects is the subject of a controversial discussion and is beyond the scope of this paper. Nevertheless, the mere similarity of these magneto-resistance effects suggests that the charge carrier localization in our [LSCO15\_1/LCO\_7]$\rm{^A}$ bilayer is also somehow related to a slowing down and freezing of the Cu$^{2+}$ spins into a domain state with a short-range antiferromagnetic order and a glassy dynamics.
\begin{figure}[!htbp] % Figure10: RT of  LSCO06-BL
\centering
\includegraphics{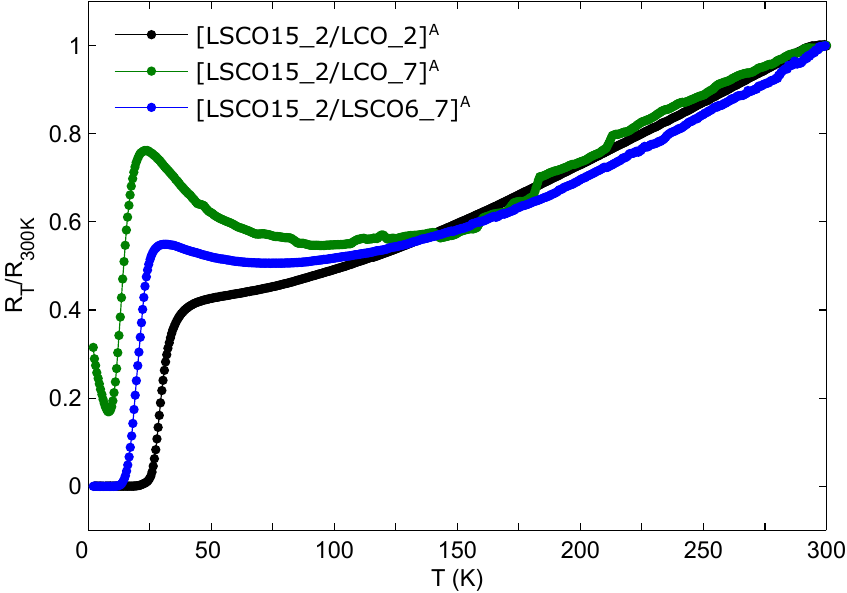}
\caption{\label{fig-LSCO06} R-T curves normalized to the value at 300\,K of the bilayers [LSCO15\_2/LCO\_2]$\rm{^A}$, [LSCO15\_2/LCO\_7]$\rm{^A}$ and [LSCO15\_2/LSCO6\_7]$\rm{^A}$. }
\end{figure}
\subsection{Origin of charge carrier localization}\label{origin}
\begin{figure}[ht] % Figure11: RT of  LSCO-LCMO
\centering
\includegraphics{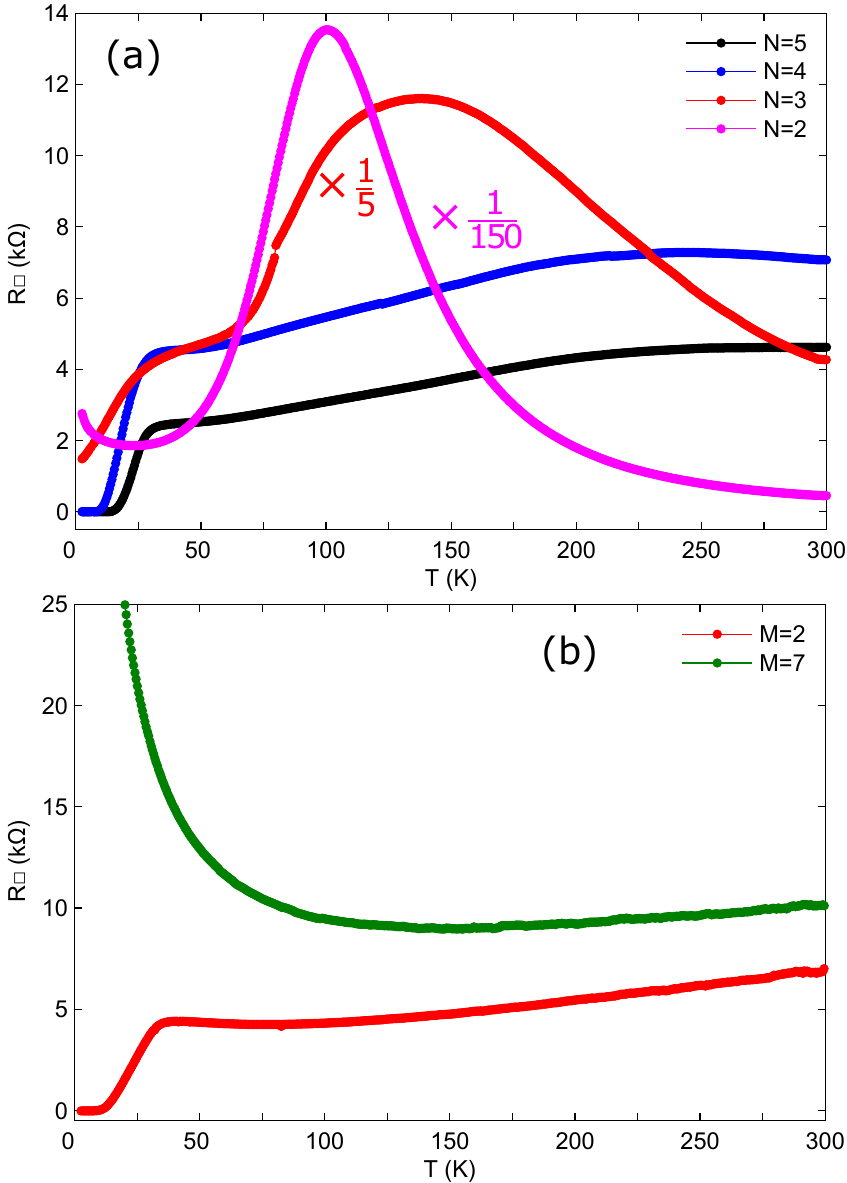}
\caption{\label{fig-lsco-lcmo} (a) {\rs}-T curves for a series of [LSCO15\_N/LCMO\_20]$\rm{^B}$ bilayers. The dome-shaped features in the {\rs}-T curves for N=2 and 3 arise from the metal-to-insulator transition of the ferromagnetic LCMO layer. (b) {\rs}-T curves of [LSCO15\_3/LCO\_M]$\rm{^B}$ bilayers.}
\end{figure}

This raises the question of why this kind of spin freezing and the related localization of the charge carriers at the LSCO15/LCO interface is so strongly dependent on the thickness of the LCO capping layer. There are at least three mechanisms that we can think of. These are (i) a transfer of a rather large amount of holes between the LSCO15 and LCO layers which leads to a strongly underdoped state in both LSCO15 and LCO, (ii) a strain-effect that is imposed on the LSCO15 layer due to the LCO top layer which has a different intrinsic lattice constant and a larger orthorhombic splitting, and (iii) an AF proximity effect due to the long range AF order of the LCO layer. {\par}
As to point (i), a strongly underdoped state in the LSCO15 layer may arise either from oxygen vacancies in the LSCO15 layer, from the inter-diffusion of Sr ions from LSCO15 to LCO, or else from a massive transfer of holes into the LCO layer. {\par}
One may indeed suspect that the M=7 LCO layer with a thickness of about 9\,nm is blocking the oxygen diffusion into the LSCO15 layer during the \textit{in situ} annealing of the bilayer. However, this is not the case since we have verified that an extended duration of the \textit{in situ} annealing procedure does not have any significant influence on the R-T curves of the [LSCO15\_N/LCO\_7]$\rm{^A}$ bilayers shown in Fig.~\ref{fig-rt-thickLCO}. Furthermore, this oxygen vacancy scenario is also not consistent with our results of a [LSCO15\_2/LSCO6\_7]$\rm{^A}$ bilayer. As shown in Fig.~\ref{fig-LSCO06}, its R-T curve reveals a metallic behavior in the normal state and a rather sharp and complete superconducting transition with of $\rm{T_{c,on}}{\approx}$30\,K and $\rm{T_{c}}{\approx}$13\,K which are in contrast with the strong resistivity upturn toward low temperature and the incomplete superconducting transition of the corresponding [LSCO15\_2/LCO\_7]$\rm{^A}$ bilayer.  This is despite the fact that the topmost $\mathrm{La_{1.94}Sr_{0.06}CuO_4}$ (LSCO6) layer has virtually the same thickness as the corresponding LCO layer with M=7. {\par}
The scenario of an inter-diffusion of a rather large amount of Sr ions into the LCO layer is also unlikely. Such an intermixing would be mainly confined to the interface region, and thus should not depend on the thickness of the LCO layer that is grown on top of LSCO15. Accordingly, the detrimental effect on the metallic and superconducting response of the LSCO15 layer should be seen for the bilayer with M=2 as much as for the one with M=7. {\par}
This leaves the scenario of a long-range transfer of holes from LSCO15 to LCO. A complete delocalization of the holes over the entire [LSCO15\_1/LCO\_7]$\rm{^A}$ bilayer would indeed result in an average doping of about 0.02 holes per $\rm{CuO_2}$ plane, and thus could explain the charge localization and the complete suppression of superconductivity at low temperature~\cite{Ando2002}. However, this scenario is in contradiction with the conclusions of Ref.~\cite{Smadici2009} where it was shown for a superlattice consisting of LCO and $\rm{La_{1.64}Sr_{0.36}CuO_4}$ that the transfer of holes from $\rm{La_{1.64}Sr_{0.36}CuO_4}$ to LCO involves a length scale of only about 6\,{\AA} which is less than one unit cell. Moreover, it has been reported in Ref.~\cite{Wu2013} that the chemical potentials of undoped LCO and optimally doped LSCO15 are almost equal such that no charge transfer is expected across the LSCO15/LCO interface. On the other hand, such a substantial charge transfer of about 0.2 holes per interfacial Cu ion has been reported to occur at the interface between a cuprate high-$\rm{T_c}$ superconductor like LSCO15 and the half-metallic ferromagnet $\rm{La_{2/3}Sr_{1/3}MnO_3}$ (LSMO)~\cite{DeLuca2014}. Nevertheless, the {\rs}-T curves of a series of [LSCO15\_N/LCMO\_20]$\rm{^B}$ bilayers in Fig.~\ref{fig-lsco-lcmo}(a) do not exhibit such steep upturn of the resistivity toward low temperature as it is seen for a [LSCO15\_3/LCO\_7]$\rm{^B}$ bilayer in Fig.~\ref{fig-lsco-lcmo}(b). For the [LSCO15\_N/LCMO\_20]$\rm{^B}$ bilayers there is merely a suppression of the $\rm{T_c}$ values as compared to the [LSCO15\_N/LCO\_2]$\rm{^B}$ bilayers in Fig.~\ref{fig-rt-bl}(b). The latter effect is expected due to pair breaking from the exchange coupling with the ferromagnet LCMO~\cite{Soltan2005}. These observations make it rather unlikely that a long-range charge transfer is responsible for the strong localization effect at the LSCO15/LCO interface. {\par}
With respect to point (ii), it has been shown that a complete suppression of superconductivity and a localization of the charge carriers occur in $\mathrm{La_{1.8-x}Eu_{0.2}Sr_xCuO_4}$ single crystals with x$<$0.2~\cite{Klauss2000}. In this system the smaller size of the $\rm{Eu^{3+}}$ ions, as compared to $\rm{La^{3+}}$, gives rise to a larger tilting of the $\rm{CuO_6}$ octahedra and a structural phase transition from the low-temperature-orthorhombic (LTO) to a low-temperature-tetragonal (LTT) phase~\cite{Klauss2000}. It has been shown in Ref.~\cite{Buchner1994} that the rotation of the $\rm{CuO_6}$ octahedra is a control parameter. At a critical angle of about 3.6$^{\circ}$, there is a transition from the superconducting state to an insulating state with a static AF stripe order. The latter persists up to a doping level of x$<$0.2. A similar structural effect, like a tilting of the $\rm{CuO_6}$ octahedra, may also occur in our LSCO15/LCO bilayers with M=7. For example, it may arise from a strain gradient in the LSCO15 layer which is clamped between the substrate with a lattice mismatch of -0.5\,\% (compressive strain) and the LCO layer with a lattice mismatch of +0.7\,\% (tensile strain). This strain gradient may also result in misfit dislocations in the ultrathin LSCO15 layer. These can also lead to a suppression of superconductivity and a charge localization since they tend to increase the scattering rate of the mobile holes and pin the short-range spin and charge order~\cite{Stilp2013}. This strain gradient should indeed depend on the thickness of the LCO layer and thus remains a viable option. {\par}
Finally, as to point (iii), there is the possibility of an AF proximity effect due to the long range AF order of LCO. It was previously reported that this static AF order occurs only if the thickness of the LCO layer exceeds a threshold value of about 2.5\,UCs~\cite{Suter2011}. The absence of the AF order in the very thin LCO layers has been explained in terms of quantum fluctuations which are enhanced due to the reduced dimensionality. Accordingly, a static AF order in the LCO layer should only occur for the bilayer with M=7, but not for the one with M=2. Correspondingly, the AF exchange coupling across the LSCO15/LCO interface should be much stronger for M=7 than for M=2. In agreement with this conjecture, the R-T curves in Fig.~\ref{fig-LSCO06} show a stronger resistive upturn and suppression of superconductivity in the [LSCO15\_2/LCO\_7]$\rm{^A}$ bilayer than in the [LSCO15\_2/LSCO6\_7]$\rm{^A}$ bilayer for which the LSCO6 layer reportedly has only weak, short-ranged AF correlations~\cite{Kastner1998}. This AF proximity effect, albeit it is expected to be fairly weak~\cite{Hucker2004}, can enhance the AF correlations and slow down the corresponding fluctuations in the adjacent optimally doped LSCO15 layer. Since it is well known that the charge dynamics is strongly affected by the AF correlations, a slowing down and eventual freezing of the AF fluctuations in the vicinity of the LSCO15/LCO interface may explain that the charge localization occurs at a substantially higher hole concentration than in the bulk. {\par}
This leaves us either with the scenario of a structural distortion in the LSCO15 layer that is induced or enhanced by the thicker LCO top layer or, likewise, of an AF proximity effect due to the long-range AF order in the LCO that leads to a slowing down and freezing of the spin fluctuations in the adjacent LSCO15 layer. Further experiments that are beyond the scope of this paper will be required to identify the relevant mechanism.
\section{Summary}
In summary, we have reported the pulsed laser deposition (PLD) growth of ultrathin {\oplsco}/{\lco} bilayers with a controlled number of N unit cells (UCs) of {\oplsco} (LSCO15) and M\,UCs of {\lco} (LCO) that was monitored with \textit{in situ} RHEED. Notably, we found that the gas environment in the PLD chamber has a decisive role on the growth mode of the first unit cell of LSCO15 (or the first two LSCO15 monolayers) next to the SLAO substrate. Especially the first monolayer has a significant amount of disorder if the growth is performed in an atmosphere of pure $\rm{N_2O}$ gas (growth type-B), whereas this disorder is strongly reduced when growing in a mixture of $\rm{N_2O}$ and $\rm{O_2}$ gas (growth type-A). For both growth types, the XRD studies confirmed the epitaxial growth of the samples with the $c$-axis along the surface normal. The reciprocal space maps (RSMs) showed that a [LSCO15\_2/LCO\_2]$\rm{^A}$ bilayer is fully strained, while a partial strain relaxation occurs for a [LSCO15\_2/LCO\_7]$\rm{^A}$ bilayer. Remarkably, the ultrathin LSCO15 layers of growth type-A are superconducting down to a thickness of one unit cell. For the corresponding bilayers prepared with growth type-B there exists an offset of one unit cell in the thickness of the LSCO15 layer that is required to obtain a signature of superconductivity. A surprising finding is that the conducting and superconducting properties of LSCO15 are also strongly dependent on the thickness of the LCO capping layer. The deposition of a LCO capping layer with a thickness of 7\,UCs leads to a strong localization of the charge carrier in the adjacent LSCO15\,UC and fully suppresses its superconducting response. The magneto-transport data point toward a magnetic origin of this charge carrier localization since they resemble the ones reported for weakly hole doped LSCO single crystals that are in the so-called {\lq{cluster-spin-glass state}\rq}. We discussed several mechanisms that could explain this kind of charge localization in an ultrathin LSCO15 layer that is adjacent to a thick LCO layer.

\begin{acknowledgments}
The work at the University of Fribourg has been supported by the Schweizer Nationalfonds (SNF) through grants No. 200020-153660, 200020-172611 and CRSII2-154410/1. Research at Universidad Complutense de Madrid has been sponsored by the Spanish  MINECO grant MAT2015-66888-C3-3-R and the European Research Council PoC MAGTOOLS \#713251.
\end{acknowledgments}

%% Appendix
\begin{appendix}
\section{2D variable-range hopping conduction in [LSCO15\_1/LCO\_7]$\rm{^A}$}{\label{app-vrh}}
When charge carriers localize in the background of a random potential, the charge transport is often well described by the variable-range hopping (VRH) model~\cite{Mott1969,Mott1978} that yields the following expression for the T-dependence of the resistance:
\begin{equation}\label{eq-vrh}
\mathrm{{\ln}R={\ln}R_0+{({\frac{T_0}{T}})}^{\frac{1}{d+1}}}
\end{equation}
\begin{figure}[!htbp] % Figure12: VRH fitting
\centering
\includegraphics{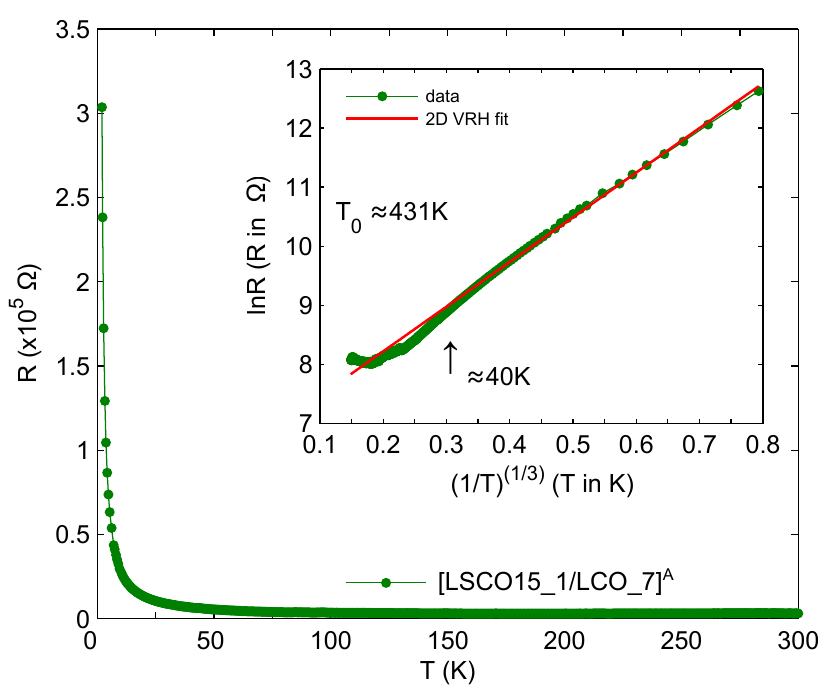}
\caption{\label{fig-vrh} R-T curve of the bilayer [LSCO15\_1/LCO\_7]$\rm{^A}$. Inset: Best fit to the low temperature data with the 2D VRH model (red solid line).}
\end{figure}
where $\rm{R_0}$ is a constant, $\rm{T_0}$ is a characteristic temperature and d is the dimensionality of the system. The localization length, $\xi$, is inversely proportional to $\rm{T_0}$. Figure~\ref{fig-vrh} shows the R-T curve of the bilayer [LSCO15\_1/LCO\_7]$\rm{^A}$ (symbols) together with a fit using the 2D (d=2) VRH model ( red solid line). The model reproduces the low-temperature data very well in the range below about 40\,K. This suggests that the charge carriers are localized below this temperature and the conduction occurs via hopping processes. It was previously reported for 100\,nm thick LSCO films with x=0.03 and 0.05 that the low-temperature conduction mechanism follows such a 2D VRH model~\cite{Shi2013}. At these low doping levels, the few holes are embedded in a background of an inhomogeneous short-range antiferromagnetic order of the $\rm{Cu^{2+}}$ spins which form a so-called {\lq{cluster-spin-glass}\rq}~\cite{Kastner1998,Shi2013}.
\end{appendix}
% The bibliography
%\bibliography{biblio}
%
\end{document}